\documentclass[traditabstract]{aa}
\usepackage{epsfig,graphicx}
\usepackage{natbib}
\bibpunct{(}{)}{;}{a}{}{,}    

\begin{document}

\title{Basal magnetic flux and the local solar dynamo}
\titlerunning{Basal magnetic flux and the local solar dynamo}
\author{J.O. Stenflo\inst{1,2}}
\institute{Institute of Astronomy, ETH Zurich, CH-8093 Zurich \and Istituto Ricerche Solari Locarno, Via Patocchi, CH-6605 Locarno Monti, Switzerland}

\date{}

\abstract{
The average unsigned magnetic flux density in magnetograms of the
quiet Sun is
generally dominated by instrumental noise. Due to the entirely
different scaling behavior of the noise and the solar magnetic
pattern it has been possible to determine the standard
deviation of the Gaussian noise distribution and remove the noise
contribution from the average unsigned flux density for the whole
15-yr SOHO/MDI data set and for a selection of SDO/HMI
magnetograms. There is a very close correlation between the MDI 
disk-averaged unsigned vertical flux density and the sunspot number, and
regression analysis gives a residual level of 2.7\,G when the sunspot
number is zero. The selected set of HMI magnetograms, which spans the most quiet
phase of solar activity, has a lower limit of 3.0\,G to the
noise-corrected average flux density. These apparently
cycle-independent levels may be identified 
as a basal flux density, which represents an upper limit to the possible
flux contribution from a local dynamo, but not evidence for its
existence. The 3.0\,G HMI level, when scaled to the Hinode spatial 
resolution, translates to 3.5\,G, which means that the much
higher average flux densities always found by Hinode in quiet regions
do not originate from a local dynamo. The contributions to the
average unsigned flux density come almost exclusively from the
extended wings of the probability density function (PDF), also in the case of HMI
magnetograms with only basal-level magnetic flux. These wings 
represent intermittent magnetic flux. As the global 
dynamo continually feeds flux into the small scales at a fast rate through turbulent
shredding, a hypothetical local dynamo may only be relevant to the
Sun if its rate of flux build-up can be competitive. While the global 
dynamo appears to dominate the magnetic energy spectrum at all the
resolved spatial scales, 
there are indications from the observed Hanle depolarization in atomic 
lines that the local dynamo may dominate the spectrum
at scales of order 1-10\,km and below. 
}

\maketitle

\section{Introduction}\label{sec:intro}
Dynamo processes are responsible for the generation of macroscopic 
magnetic fields from feeble seed fields \citep{stenflo-larmor1919,stenflo-elsasser1946,stenflo-elsasser1956}. The
solar dynamo, which is the source of the 11-year activity cycle with
its sunspots, flares, CMEs, prominences, etc., is a prototype of a
cosmic dynamo. The basic  ingredients of the Sun's dynamo are the same
as for the galactic and planetary dynamos: interactions between magnetic fields and
turbulence in an electrically conducting and rotating medium. The
rotation breaks the left-right symmetry of the turbulence through the
Coriolis force, thereby generating a large-scale net helicity that is
the source of the dynamo-produced magnetic field
\citep{stenflo-parker55,stenflo-steenkrause69}.  

While the symmetry breaking mechanism can produce magnetic patterns on
large scales, turbulent shredding ensures that the magnetic scale 
spectrum extends over many orders of
magnitudes, down to the magnetic 
diffusion limit near 10-100\,m, far below the observationally resolved
scales on the Sun \citep[cf.][]{stenflo-s12aa1}. The magnetic energy that is injected at large
scales by the dynamo quickly cascades all the way down the scale 
spectrum. The turbulent scale redistribution of the globally
generated magnetic energy gives the Sun's magnetic
field a fractal-like appearance. 

In the context of numerical simulations of magneto-convection the
concept of a small-scale, ``local dynamo''  that operates near the
solar surface has often been used to account for much
of the small-scale magnetic structuring
\citep{stenflo-kazantsev68,stenflo-petrovay93,stenflo-cattaneo99,stenflo-brandsub05}. In contrast to the previously
discussed ``global dynamo'', the local dynamo does not need a rotating
medium or the breaking of the left-right symmetry of the convective motions but
simply depends on the way convection tangles the magnetic
field. However, in the great majority of the numerical simulations done
to date, the amount of magnetic energy that is generated at small
scales depends on the initial conditions of the simulation, and
therefore they do not represent any local dynamo. 

A local dynamo only exists in the simulations if, without input from any 
global dynamo, magnetic energy can be built up from
an initial seed field, and if the results of the simulations are
independent of the choice of seed field. One clear demonstration that such a local
dynamo is possible on the Sun is that of \citet{stenflo-vs07}, but, as
the authors point out, their results depend on chosen simulation
idealizations, like the subgrid model for the viscosity. Therefore
this demonstration 
cannot be interpreted as proving that a local dynamo contributes
significantly to the observed small-scale magnetic fields on the real Sun. 

The question how relevant a local dynamo is for the Sun cannot be determined by numerical
simulations alone, because their results depend on the 
idealizations and simplifications that have to be introduced. Instead
we have to look for empirical answers derived from the observed
properties of the Sun, trying to separate the relative
contributions from the global and local dynamos. 

We know that the cyclic behavior of the Sun's magnetic field (22-yr
magnetic cycle with the 11-yr sunspot cycle) is the result of global
dynamo action. In contrast the local dynamo is related to the quiet
Sun and produces magnetic fields that are statistically time
invariant. Therefore a local dynamo can only contribute to a possible
``basal flux'', a minimum magnetic flux level 
that is always present everywhere and every time on the Sun, even in
the complete absence of sunspots and solar activity. The
observed presence of a basal X-ray flux for slow-rotating cool stars
\citep{stenflo-schrijver87} has been interpreted as empirical evidence for the
existence of dynamo action in non-rotating plasmas without symmetry
breaking from the Coriolis effect \citep{stenflo-bercikfisher05}. 

In the next sections we will use data from SOHO/MDI
\citep{stenflo-scherreretal95} and SDO/HMI
\citep{stenflo-scherrer12,stenflo-schouetal12} to  
determine the Sun's basal magnetic flux density level. Since the apparent,
time-invariant flux level has its main contribution from instrumental
noise, we have developed a technique to statistically separate the
noise and solar contributions from each other, to find the noise-free,
intrinsic basal flux level. Such a separation is possible because the
noise and the solar field obey completely different scaling laws. The
basal flux density level is resolution dependent, but in a way that is well
defined from a known scaling law. The low value of the determined
basal flux density allows us to place a tight limit to the possible contribution of
a local dynamo to the observed small-scale magnetic fields on the Sun.


\section{Correlation between unsigned flux and sunspot number}
Recently a global analysis of the complete set of 96 minute cadence
SOHO/MDI full disk magnetograms, 73,838 of them covering the 15-yr 
period May 1996 -- April 2011, was carried out with the aim of
exploring the properties of bipolar magnetic regions
\citep{stenflo-sk12}.  As a byproduct of this analysis we determined 
for each magnetogram the average value $B_{\rm ave}$ of the unsigned
vertical flux density $\vert B_v\vert$ over a circular region around
disk center, for which the normalized radius vector $r/r_\odot$ was
less than 0.1, 0.2, $\ldots$, 0.8, and 0.9, respectively, 
  including all pixels in the average (regardless of whether they
  occur in sunspots or not). This allows 
us to explore how well the value of $B_{\rm ave}$ correlates with the sunspot
number for the various sizes of the averaging regions. 

The vertical flux density $B_v$ was obtained from the line-of-sight
component $B_\parallel$ of the magnetograms through the
simple relation $B_v = B_\parallel /\mu$, where $\mu$ is the cosine of
the heliocentric angle of the respective pixel in the image. This
procedure is based on the assumption that the fields are on average
oriented in the vertical direction. While this is the only practically
feasible assumption that can be made, it is physically justified (at
least for statistical purposes outside active regions) by the circumstance that most ($>
90$\,\%) of the net flux recorded with the MDI 4\,arcsec resolution comes
from collapsed, kG type flux elements
\citep{stenflo-hs72,stenflo-s73,stenflo-book94}. Such strong
  fields get pushed in the vertical direction by the 
buoyancy forces in the photospheric layers where the fields are
measured. Higher (above the observed layers)
the field starts to become force-free and may  increasingly deviate from the
vertical. Errors due to this assumption will contribute to statistical
scatter. The circumstance that the relation between the disk averaged
flux density and the sunspot number turns out to be so tight, as we
will see next, lends support to this procedure. 

\begin{figure}
\resizebox{\hsize}{!}{\includegraphics{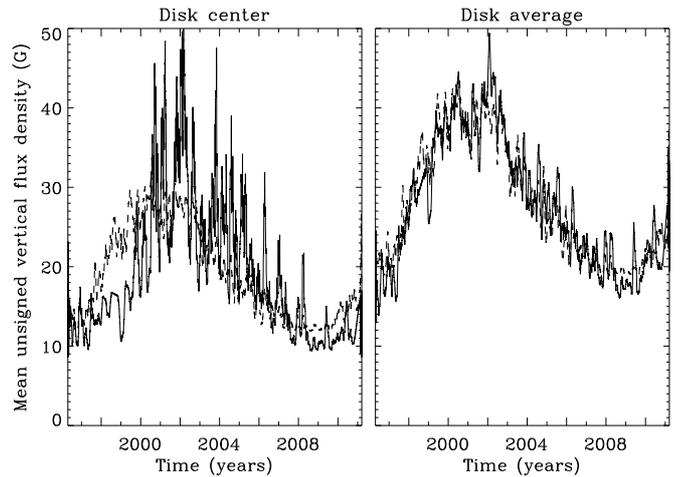}}
\caption{Time series of the average of the unsigned vertical flux
  density, smoothed with a 1-month wide time window (solid
  lines). Left panel: averaging over the disk region $r/r_\odot
  <0.1$. Right  panel: averaging over the disk region $r/r_\odot
  <0.9$. The dashed curves use a second-order polynomial in the
  sunspot number (also smoothed with a 1-month time window) to fit the
  solid curves. The fit gets tighter as the disk averaging area
  increases. 
}\label{fig:sunspotcorr}
\end{figure}

Figure \ref{fig:sunspotcorr} shows as the solid curves the value of
$B_{\rm ave}$, time smoothed with a 1-month wide window, for two
choices of the averaging region: $r/r_\odot <0.1$ (left panel), and
$r/r_\odot <0.9$ (right panel). The dashed curves represent the
monthly averages of the international sunspot number $R_z$, converted
to a magnetic field scale $B_{R_z}$ through the second-order relation 
\begin{equation}
B_{R_z}=b_0 +b_1 R_z +b_2 R_z^2\,.\label{eq:brz}
\end{equation}
This relation is fitted to the solid curves with the coefficients as
the three free fit parameters, which gives us the dashed curves. 

We see from Fig.~\ref{fig:sunspotcorr} that the disk-averaged value of
$\vert B_v\vert$ (right panel) can be nearly perfectly modeled, the
fit is surprisingly tight, implying that there is an almost 
one-to-one relation between $B_{\rm ave}$ and the sunspot number. If we
however only use the innermost portion of the solar disk for the
averaging of $\vert B_v\vert$ (left panel), then $B_{\rm ave}$
fluctuates rather wildly, in particular during times of high solar
activity, and the fit with the sunspot number is poor. The reason is
of course that the magnetic field in the disk center region is not
representative of the overall level of magnetic activity on the
Sun. The presence or absence of individual low-latitude active regions
govern the behavior of $\vert B_v\vert$ near disk center. One needs to
account for the cumulative magnetic contributions over a large portion
of the disk to get a good representation of solar activity. 

It is nevertheless surprising that the sunspot number $R_z$ turns out
to be such a nearly perfect index for the overall magnetic activity of
the Sun as represented by the disk average of the unsigned vertical
flux density. The largest deviations between the
model fit and the data in the right panel of
Fig.~\ref{fig:sunspotcorr}  occur in regions of significant data gaps
in the MDI time series (1998.48 - 1998.89 and 1998.97 - 1999.17), so
the physically relevant fit is even better than it may appear in 
the plot. As the size of the averaging region is increased, the 
fit between the $R_z$ model and $B_{\rm ave}$ becomes quite 
tight already for $r/r_\odot <0.5$, and the fit improves only slowly 
as we go to still larger averaging areas. For $r/r_\odot <0.9$ (right panel
of Fig.~\ref{fig:sunspotcorr}) the fit parameters of the model have
the values (in G) $b_0 =19.0$, $b_1 =0.24$, and $b_2 =-5.5\times
10^{-4}$.  

Coefficient $b_0$, as defined by Eq.~(\ref{eq:brz}), can be interpreted
as representing the apparent basal flux density, the
minimum level reached in the absence of sunspots. However,
instrumental noise contributes to 
a fictitious non-zero basal level, since the noise does not average
out to zero when we average over the absolute, unsigned flux
densities. Recently a careful noise analysis of the MDI magnetograms
by \citet{stenflo-liuetal12} showed that the instrumental noise is
indeed of the same order as our apparent basal flux density level of
19\,G. This indicates that the intrinsic level of the basal flux
density must be much smaller. A determination of its actual value
requires a very careful separation and removal of the noise
contribution. The procedure for doing this is described in the next
section.


\section{Statistical removal of MDI noise}\label{sec:mdinoise}
Fortunately it is possible to statistically separate the noise
contribution from the intrinsic solar contribution, since the Gaussian
noise and the fractal-like solar magnetic pattern obey entirely
different scaling laws. The two contributions however combine in a
way that is neither linear nor quadratic. Before we can model the
scaling behavior we need to determine the form of the non-linear
relation between the noise-affected and noise-free average unsigned
flux densities, which will be done in the next subsection. 


\subsection{Convolution of noise with the intrinsic
  PDF}\label{sec:conv}
Let $P(B_v)$ be the area normalized PDF (probability density function)
for the intrinsic, noise-free flux densities. Then the average
unsigned flux density is 
\begin{equation}
B_{\rm ave} =\int P(B_v)\,\vert B_v\vert\,{\rm d}B_v\,.\label{eq:bavepdf}
\end{equation}
As shown in a detailed analysis of Hinode SOT/SP data for the quiet
Sun disk center \citep{stenflo-s10aa}, $P(B_v)$ is characterized by an
extremely narrow core peak centered at $B_v=0$, which can be modeled
by a stretched exponential. This peak is surrounded
by quadratically declining damping wings that extend out to the kG
region. The contribution to $B_{\rm ave}$ is completely dominated by
the contribution from the damping wings, since the inner PDF core is
so narrow. 

The apparent, noise-affected PDF, $P_{\rm app}$, is obtained from the
intrinsic $P$ through direct convolution with the noise distribution,
which can safely be assumed to have a Gaussian shape. For a Gaussian
distribution with standard deviation $\sigma$, the average of the
unsigned field value of the distribution is $0.798\,\sigma$. If the
intrinsic PDF were also Gaussian, then the convolved distribution
would remain Gaussian with a sigma that is obtained by adding the two
individual sigmas quadratically. In contrast, when convolving two
Lorentzian profiles (which have quadratically declining wings, like
the solar PDFs), their half widths add linearly. In the present
case, however, we are convolving a Gaussian with a function that is
more Lorentzian like. We would therefore expect that the average
unsigned field values would combine in a way that is neither
quadratic nor linear, but somewhere in between. In the following we
will determine the form of this relation. 

\begin{figure}
\resizebox{\hsize}{!}{\includegraphics{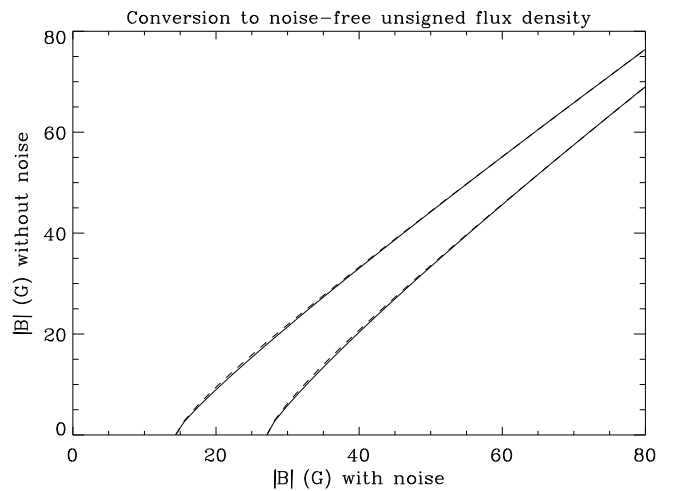}}
\caption{Relation between the noise-free average unsigned vertical
  flux density (vertical axis, in the text denoted $B_{\rm ave}$) and
  the corresponding noise-affected flux density (horizontal axis, in
  the text denoted $B_{\rm app}$), for two values of $\sigma$ for the
  noise distribution, 18 and 34\,G. The solid curves are obtained
  through numerical convolution of PDFs that are representative of quiet-sun magnetic
  fields (as derived from Hinode analysis) with Gaussian noise
  distributions, while the dashed curves are analytical
  representations of the solid curves in terms of Eqs.~(\ref{eq:noisemod}) and
  (\ref{eq:alpha}). 
}\label{fig:bubun}
\end{figure}

The noise-affected average unsigned flux density
$B_{\rm app}$, which is obtained from Eq.~(\ref{eq:bavepdf}) if we
replace  $P$ by $P_{\rm app}$, can be determined directly from any
given value for the noise $\sigma$ by direct numerical Gaussian
convolution of $P$. To obtain this relation also as a function of
$B_{\rm ave}$, representing solar regions with varying amount of
magnetic flux, we have used the analytical shape of the symmetrized
intrinsic PDF that represented our Hinode disk center data set
\citep{stenflo-s10aa}, and produced a family of $P(B_v)$ distributions
by varying the value of the damping parameter that governs the
strength of the damping wings of the PDF and therefore also the value
of  $B_{\rm ave}$. The result of these numerical convolutions and
integrations are shown by the solid curves in Fig.~\ref{fig:bubun},
computed for two chosen values of the noise $\sigma$, 18 and
34\,G. Note that the average unsigned field values for these two
Gaussians are smaller than these sigmas by the factor 0.798 and are
thus 14.4 and 27.1\,G, respectively. 

Next we have replaced the Hinode-type shape of $P$ with a Lorentzian
function and repeated the calculations. Convolution of a Lorentzian
function with a Gaussian gives the well-known Voigt function. The resulting curves are
practically indistinguishable from the solid curves in 
Fig.~\ref{fig:bubun} and are therefore not plotted. This result is not
unexpected, since the non-Lorentzian core of the Hinode PDF $P$ does
not contribute significantly to the value of $B_{\rm ave}$, and the
PDF wings have the same shape as a Lorentzian function. Therefore
our modeling is insensitive to the true shape of the PDF core region. 

As we will later use inversion techniques to determine the actual noise
in the magnetograms and calculate the scaling law for the
noise-affected flux density, we need
to represent the numerical results in Fig.~\ref{fig:bubun} (the solid
curves) in an analytical form, valid for arbitrary choices of the
noise $\sigma$. Based on the considerations that we have made before
about the way in which Gaussian and
Lorentzian functions combine, our analytical model for the conversion
of the noise-affected $B_{\rm app}$ to the noise-free $B_{\rm ave}$
for a given value of the standard deviation $\sigma$ of the noise
distribution is 
\begin{equation}
B_{\rm ave} =[\,B_{\rm app}^\alpha\,-\,(0.798\,\sigma)^\alpha\,]^{(1/\alpha)}\,.\label{eq:noisemod}
\end{equation}
For two Gaussian functions $\alpha$ would be 2.0, for two Lorentzian
functions it would be 1.0. By choosing $\alpha$ (through trial and error)
to be 
\begin{equation}
\alpha=1.36-0.004\,\sigma+0.0034\, B_{\rm app}\,,\label{eq:alpha}
\end{equation}
we obtain the dashed curves in Fig.~\ref{fig:bubun}, which can be seen
to be nearly perfect representations of the exact, numerical
relations. More importantly, our analytical model remains an excellent
representation not just for the two chosen sigmas of the
two curves in  Fig.~\ref{fig:bubun}, but for arbitrary values of
$\sigma$ across the range spanned by these two $\sigma$ values. This
is the range within which we will find the actual MDI noise values to
lie. 

We note in Eq.~(\ref{eq:alpha}) that $\alpha$ falls between the
values 2 and 1 for the Gaussian and Lorentzian cases, as expected.


\subsection{Scaling law for the average unsigned flux density}\label{sec:canc}
The unsigned vertical flux density is a resolution dependent quantity
because the pattern of solar magnetic fields contains mixed magnetic
polarities across the whole range of spatial scales (resolved and
unresolved). Due to cancellation of the positive and negative 
contributions of the opposite-polartiy magnetic fluxes within each
resolution element, the net flux or apparent flux density (as averaged
over the resolution element) is reduced. As we increase the spatial
resolution more of the mixed polarity elements get 
resolved, which reduces the cancellation effects. This causes the
average of the unsigned vertical flux density to increase. 

The scaling law that governs the variation of the average unsigned
vertical flux density is called the {\it cancellation function}. It
was introduced by \citet{stenflo-pietarila09} for analysis of a
Hinode SOT/SP data set for the quiet Sun disk center, which had been
converted to vertical flux densities by 
\citet{stenflo-litesetal08}. The cancellation function
\begin{equation}
B_{\rm ave} \,\sim\, d^{-\kappa}\label{eq:canc}
\end{equation}
describes how the average unsigned flux density $B_{\rm ave}$ for a
given region on the Sun depends on the chosen resolution scale $d$.  
Through numerical smoothing of the Hinode data
\citet{stenflo-pietarila09} determined the value of the {\it
  cancellation exponent} $\kappa$ to be 0.26. The determination is
however sensitive to the influence of instrumental noise and the way
in which the polarimetric data are converted to flux densities. An
independent determination of the cancellation exponent from quiet-sun
Hinode data led to $\kappa=0.13$, a value half as large
\citep{stenflo-s11aa}. If, as it seems, the pattern of solar magnetic
fields behaves like a fractal, we expect $\kappa$ to be independent of
scale size $d$ over several orders of magnitude, including
scales larger than the MDI resolution and unresolved small scales
beyond the Hinode resolution. 

Random instrumental noise, which is spatially uncorrelated from pixel
to pixel, scales with $1/\sqrt{N}$, where $N$ is the number of pixels
inside the smoothing window (simulated resolution element), implying a
$\kappa$ of unity, representing a much steeper scaling law than that
of the magnetic-field pattern. Noise has the 
effect of spuriously steepening the cancellation function. 

The 96 minute cadence SOHO/MDI data set of full disk magnetograms that
we have used contains a mixture of magnetograms based on two distinctly different
integration times, 1\,min and 5\,min. A keyword in the fits file
header identifies which integration time was used for the given
magnetogram. Since the noise characteristics are distinctly different
for the two choices of integration time, we have divided the entire
data set into two subsets, each representing the respective integration
time. 

In their recent analysis of the noise in the MDI magnetograms,
\citet{stenflo-liuetal12} estimated the disk-averaged value for the standard
deviation $\sigma$ of the noise distribution to be 16.2 and 26.4\,G
for the magnetograms with 5-min and 1-min integrations,
respectively. They considered these values to only be upper limits,
since they had been obtained by 
fitting Gaussians to the PDF cores in quiet solar regions,
assuming that the solar contribution to the core width can be
neglected. If this assumption is incorrect, the actual noise is
smaller. 

The values of \citet{stenflo-liuetal12} are for the disk-averaged
line-of-sight component $B_\parallel$, while our analysis is based on
the vertical flux density $B_v=B_\parallel/\mu$ within radius vector
$r/r_\odot <0.9$. Since the average of $1/\mu$ within this radius
vector is 1.39, we should multiply by this factor to translate the Liu et
al. values to a scale that is comparable to ours. Their values then become 22.5
and 36.7\,G. Note however that this crude translation only represents an
estimate, because the noise is not constant over the disk 
but has a large-scale pattern. Nevertheless it serves as a
reference to check the consistency of the analysis. As we will see in
the next subsection, our use of the scaling laws to determine
the noise $\sigma$ leads to the values 18.8 and 33.2\,G, which are 
10-20\,\%\ smaller than the corresponding values of
\citet{stenflo-liuetal12}. As their results represent upper
limits, our results are fully consistent with theirs. 

\begin{figure}
\resizebox{\hsize}{!}{\includegraphics{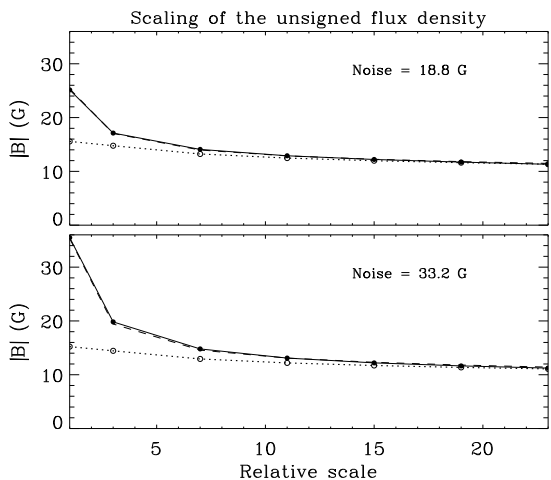}}
\caption{Average unsigned vertical flux density vs. relative size (in
  pixel units) of the smoothing window, illustrated for two
  representative magnetograms, one obtained with 5-min integration
  (upper panel), the other with 1-min integration (lower panel). The
  solid curves represent the observed, noise-affected values $B_{\rm
    app}$, the dotted curves the noise-free values $B_{\rm ave}$,
  while the dashed curves are obtained through combination of $B_{\rm
    ave}$ with the noise $\sigma$ according to the model described by
  Eqs.~(\ref{eq:noisemod}) and (\ref{eq:alpha}). 
}\label{fig:scaling}
\end{figure}

In Fig.~\ref{fig:scaling} we illustrate for two representative
magnetograms how the two scaling laws, with
$\kappa=1$ for the noise and $\kappa=0.13$ for the magnetic fields,
combine to give a scaling behavior that can be fitted to the
observational data. The average unsigned vertical flux density over
$r/r_\odot <0.9$ is plotted as a function of the relative scale $d$,
in units of the MDI pixel size. $d$ represents the side of the
square-shaped window used for numerical spatial smoothing. The solid
curves are the actually observed values $B_{\rm app}$ of the average unsigned
flux density, the dotted curves the inferred intrinsic average unsigned
flux density $B_{\rm ave}$, while the dashed curves represent the
modeled combination of $B_{\rm ave}$ with a noise $\sigma$ (for
$d=1$) of 18.8\,G (upper panel) and 33.2\,G (lower panel). We see that
the agreement between the model (dashed) and observations (solid) is
nearly perfect in both cases. It is through fitting of the model to
the observations that the noise $\sigma$ is determined, as
will be described next. This determination also leads to a
verification that $\kappa=0.13$ is indeed the correct cancellation
exponent to use.


\subsection{Determination of the noise levels}\label{sec:determnoise}
For the model fitting to determine the noise contribution to the
observed average unsigned flux densities $B_{\rm app}$ we have
randomly selected 320 MDI magnetograms evenly distributed over the
15\,yr period of MDI operation. The fits header keywords reveal that
140 of these magnetograms represent recordings with 5-min
integration, 180 with 1-min integration. The magnetograms have then
been spatially smoothed with square windows having side $d=3$, 7, 11, 15, 19, and
23 pixels. For each magnetogram there are thus 7 versions with
different degrees of smoothing. They are all converted to vertical
flux densities $B_v$ through division by $\mu$ for each pixel. 
$\vert B_v\vert$ is then averaged over the disk region $r/r_\odot
<0.9$ to give the noise-affected average unsigned vertical flux
density $B_{\rm app}$ as a function of $d$. For each
magnetogram we thus obtain 7 observables (which in Fig.~\ref{fig:scaling} are
represented by the filled circles and solid lines, for the choice of
two magnetograms). 

These 7 observables serve as input for an iterative least squares
model fitting. The
fit model is based on Eqs.~(\ref{eq:noisemod}), (\ref{eq:alpha}), and
(\ref{eq:canc}), with the $B_{\rm app}$ values for the 7 different $d$
values being the observables to be fit. For convenience we introduce
the notation $B_{\rm solar}$ to refer to 
the value of $B_{\rm ave}$ for the unsmoothed case (to distinguish it
from the more general notation $B_{\rm  ave}$, which may refer to any
smoothed case and therefore depends on $d$). 
Then the $d$-independent model parameters to be determined are $B_{\rm
  solar}$, $\sigma$, and cancellation exponent $\kappa$.  For reasons
of numerical robustness we only let $B_{\rm  solar}$ and $\sigma$ be
free model parameters, while keeping the value of $\kappa$ fixed, 
but repeat the fit procedure for a sequence of $\kappa$ values to
demonstrate that only one particular value of $\kappa$ leads to
physically acceptable results. 

Let us here mention the technical detail that $B_{\rm  solar}$, the
unsmoothed version of $B_{\rm ave}$, actually 
refers to $d=2$, since according to the sampling 
theorem there must be at least $2\times 2$ pixels per resolution
element. Therefore the resolution does not increase when we go from $d=2$ to
$d=1$. In contrast, the sampling theorem does not apply to the 
noise, which does not contain spatial structures. In
Fig.~\ref{fig:scaling} we have plotted the value of $B_{\rm  solar}$ at $d=1$
(as the left ends of the dotted lines), because it relates to the
values of $B_{\rm app}$ and $\sigma$ at $d=1$. The practical
consequences of the sampling theorem are however insignificant, since
the solar scaling is so small when going from $d=1$ to 2. Thus
$2^{\kappa}$ is only 1.09 for $\kappa=0.13$. 

\begin{figure}
\resizebox{\hsize}{!}{\includegraphics{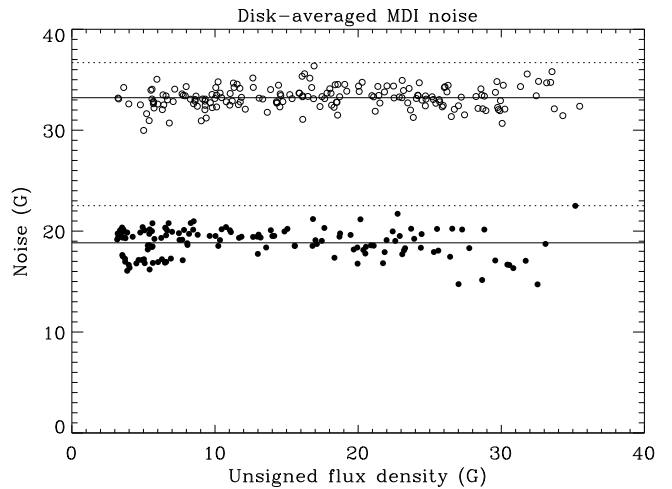}}
\caption{Standard deviation $\sigma$ of the Gaussian noise
  distribution (vertical axis) vs. noise-free average unsigned
  vertical flux density, determined from least-squares fits to the
  randomly selected 320 MDI magnetograms, assuming a cancellation
  exponent $\kappa=0.13$. The filled circles have been obtained from
  magnetograms recorded with 5-min integration, the open circles from
  magnetograms with 1-min integration. The solid horizontal lines at
  18.8 and 33.2\,G represent the average values of $\sigma$ for the
  two populations. For comparison the corresponding upper limits to
  the noise levels as derived from the analysis of
  \citet{stenflo-liuetal12} are given as the dotted lines. 
}\label{fig:fitmdi}
\end{figure}

Our least squares fitting gives us for each of the 320 magnetograms
the two determined fit parameters $B_{\rm  solar}$ and $\sigma$, the
standard deviation of the Gaussian noise 
distribution. In Fig.~\ref{fig:fitmdi} we have for a cancellation
exponent $\kappa=0.13$ plotted $\sigma$ as a
function of $B_{\rm  solar}$, using filled circles for the
magnetograms that were obtained with 5-min integrations, open circles
for the magnetograms based on 1-min integrations. We see that the
filled and open circles form two distinctly different populations, as
expected. The two horizontal lines mark the average values of $\sigma$
for each population, 18.8 and 33.2\,G, respectively. For comparison we
give as the dotted lines the upper limits  to $\sigma$, 22.5 and
36.7\,G, derived by \citet{stenflo-liuetal12} but here scaled with the
average value of $1/\mu$ inside $r/r_\odot <0.9$ to make them refer to our
vertical flux density scale. They lie 10-20\,\%\ higher than our solid
lines, which is fully consistent with the circumstance that they
represent upper limits only. 

The two magnetograms that we selected for illustration of the scaling
behavior in Fig.~\ref{fig:scaling} were chosen to have $B_{\rm
  solar}$ values in the middle of the range spanned by the 320
magnetograms. Thus the $B_{\rm  solar}$ value for the upper panel is
15.6\,G, for the lower panel 15.2\,G. The corresponding points in
Fig.~\ref{fig:fitmdi} lie almost exactly on the horizontal lines that
represent the average $\sigma$ for the respective point population. 

\begin{figure}
\resizebox{\hsize}{!}{\includegraphics{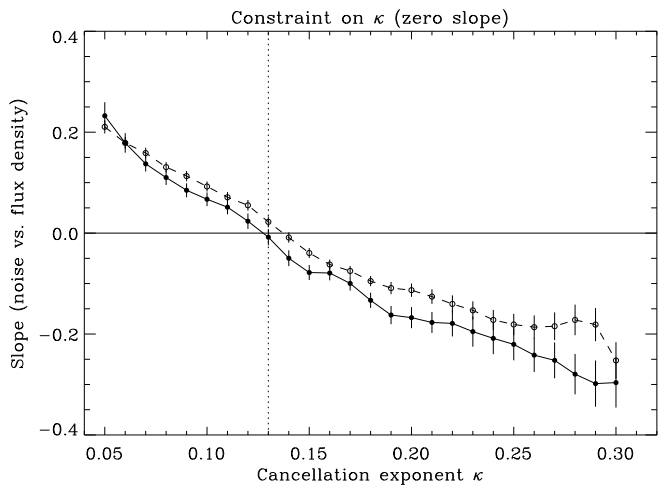}}
\caption{Regression line slopes with error bars for the point
  populations in diagrams of the type of Fig.~\ref{fig:fitmdi}, based
  on model fitting with different fixed values for the cancellation
  exponent $\kappa$. The filled circles and solid line represent
  magnetograms with 5-min integration, while the open circles and
  dashed line represent magnetograms with 1-min integration. Only zero
  slope is physically acceptable, which occurs when $\kappa=0.13$
  (marked by the vertical dotted line). 
}\label{fig:kappa}
\end{figure}

It is gratifying that there is no dependence of the fitted value of
the noise $\sigma$ on the magnetic flux level $B_{\rm  solar}$ of a
given magnetogram. Any such dependence would be unphysical if  the
noise is of instrumental origin and is fully stochastic and
  additive, like photon noise, and since the brightness-flux
  correlation is very small when recorded with 4\,arcsec resolution
  (except in sunspots, which are expected to only have a minor
  influence on the global flux averages). However, this flux
invariance of $\sigma$ turns out to be unique for 
the chosen value of cancellation exponent $\kappa$. For other values
of $\kappa$ the point distribution in $B_{\rm  solar}$ - $\sigma$
space is tilted. We have repeated the fitting procedure for a sequence
of fixed $\kappa$ values, determined the slopes of the regression line
fits to the point populations, and plotted these slopes with their
error bars as a function of $\kappa$ in Fig.~\ref{fig:kappa}. The
filled circles and solid line represent the magnetograms with 5-min
exposure, the open circles and dashed line the magnetograms with 1-min
exposure. We see that both sets of magnetograms give zero slope only
for $\kappa=0.13$, while for all other choices of $\kappa$ an
unphysical non-zero slope is found. It is this result that unambiguously
leads us to the unique value $\kappa=0.13$. 

The same value of $\kappa$ was found from an analysis of the scaling
behavior of Hinode magnetograms for the quiet-sun disk center
\citep{stenflo-s11aa}. This consistency between the widely different
Hinode and MDI data provides support for the validity of our
  noise model, while indicating that $\kappa$ is indeed invariant over a
large scale range, as expected from fractal-like behavior of the
magnetic field pattern. The circumstance that the same noise
  model with the same $\kappa$, when applied to the quite different
  HMI data set for the quiet-sun disk center, also gives a noise
  $\sigma$ that does not depend 
  on the magnitude of the unsigned flux density 
  (cf.~Fig.~\ref{fig:hminoise}), further supports the validity of our approach.

\subsection{Noise-corrected correlation with the sunspot number}\label{sec:rzcorr}

\begin{figure}
\resizebox{\hsize}{!}{\includegraphics{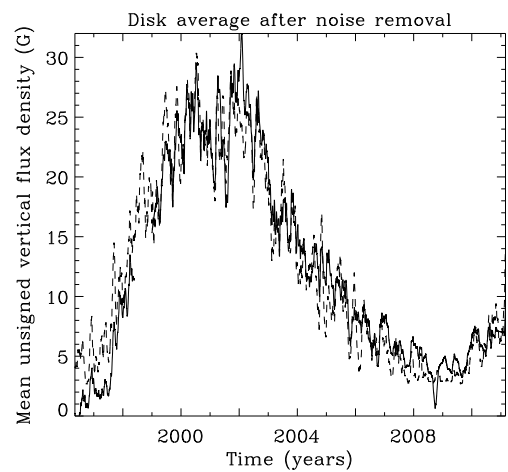}}
\caption{Time series of the noise-corrected disk average (over the disk region $r/r_\odot
  <0.9$) of the unsigned vertical flux
  density, smoothed with a 1-month wide time window (solid
  line). Due to an MDI data gap the solid line is absent in the
  interval 1998.4 - 1999.0. The dashed curve uses a second-order
  polynomial in the 
  sunspot number (also smoothed with a 1-month time window) to fit the
  solid curve. This polynomial fit gives a basal flux density (in the
  absence of sunspots) of 2.7\,G. 
}\label{fig:rzfit}
\end{figure}

Having determined the two values of the noise $\sigma$ for the 5-min
and 1-min exposure magnetograms, we can now apply Eqs.~(\ref{eq:noisemod}) and
  (\ref{eq:alpha}) to convert the noise-affected average unsigned
  vertical flux density $B_{\rm app}$ to its noise-free counterpart
  $B_{\rm ave}$. Similar to the right panel of
  Fig.~\ref{fig:sunspotcorr} we plot  as the solid curve in Fig.~\ref{fig:rzfit} $B_{\rm ave}$ for the
  set of 5-min exposure magnetograms, time-smoothed with 
  a 1-month window. Like in  
  Fig.~\ref{fig:sunspotcorr} we model this curve in terms of a
  second-order polynomial in the sunspot number $R_z$ (also
  time-smoothed with a 1-month window), defined by
  Eq.~(\ref{eq:brz}). The resulting values for the fit parameters in
  units of G are: $b_0 =2.7$, $b_1 =0.25$, and $b_2 =-5.0\times
10^{-4}$. Except for the value of $b_0$ ($B_{\rm ave}$ in the absence
of sunspots) the fit parameters ($b_1$ and $b_2$) are nearly identical
to those of Fig.~\ref{fig:sunspotcorr}. For $b_0$, however, there is
a dramatic change, from 19.0\,G in the noise-affected case to 2.7\,G
in the noise-free case. The noise-corrected value of  $b_0$ may be
interpreted as a time-invariant, basal flux density of the Sun. 

The corresponding time series of the noise-corrected $B_{\rm ave}$ 
based on the use of the 1-min 
exposure magnetograms is nearly identical although somewhat noisier 
and is therefore not illustrated separately here. The polynomial fit parameters are
slightly different: $b_0 =1.0$, $b_1 =0.28$, and $b_2 =-6.5\times
10^{-4}$. This difference gives an indication of the uncertainty in the
determination of the basal flux level. As the noise $\sigma$
for the 1-min magnetograms is almost twice as large (33.2\,G) as for
the 5-min magnetograms (18.8\,G), we give considerably more weight to
the analysis based on these lower-noise magnetograms. As we will see
below, our 2.7\,G level found with the 5-min MDI magnetograms is fully
consistent with the 3.0\,G level that we find from analysis of
SDO/HMI magnetograms. The consistency between the MDI and HMI results
supports the conclusion that the basal flux density of about 3\,G is
a real physical property of the Sun.

\section{Basal flux density from HMI data}\label{sec:basal}
The {\it Helioseismic and Magnetic Imager} (HMI) on the {\it Solar Dynamics
Observatory} (SDO) \citep{stenflo-scherrer12,stenflo-schouetal12} is
the next-generation successor of the MDI instrument on SOHO and 
has been operating since April 2010. HMI delivers line-of-sight
magnetograms with a cadence of 45\,s and a spatial resolution that
is four times higher than that of MDI, with a noise level that is
better than that of the 5-min MDI magnetograms by more than a factor of
2. This substantial improvement in quality makes the HMI data
particularly suited for a determination of the basal flux density
level. 

The advantage of the MDI data set over that of HMI is that the MDI
time series of 15\,yr covers more than a full sunspot cycle, which
allows an exploration of the correlation between disk-averaged
unsigned flux density and sunspot number, as we have done in
Figs.~\ref{fig:sunspotcorr} and \ref{fig:rzfit}. The second-order
polynomial fit to the sunspot number allowed us to identify the
constant coefficient in the polynomial fit as the basal value of the 
disk-averaged vertical flux density, since it represents the flux density
in the absence of 
sunspots. This type of approach is not possible with the HMI data,
since HMI so far only covers a small fraction of a solar cycle. 

The HMI data set on the other hand has the advantage of covering in
great detail a cycle phase when the Sun has been unusually
quiet. During this phase the average unsigned flux densities should
come closest to the minimum level that is given by the basal
flux density. 

Like with the MDI data, the apparent basal flux density
of the HMI data is dominated by the noise contribution. A 
determination of the basal flux density therefore has to be preceded
by an accurate modeling and determination of the HMI noise. This is
done in the following subsection. 

\subsection{HMI noise at disk center}\label{sec:noisedc}
For our HMI analysis we have selected 676 magnetograms, one for each
day, starting April 30, 2010. The procedure for the determination of
the standard deviation $\sigma$ of the Gaussian noise distribution
with least squares model fitting based on the different scaling
behavior of the noise and the magnetic fields is identical to the
procedure that we used for MDI and which was described in great
detail in Sect.~\ref{sec:determnoise}. For each HMI magnetogram we
thus generate 7 different versions with different degrees of spatial
smoothing. However, since we are not doing correlations with the
sunspot number, there is no reason for averaging the unsigned flux
density over most of the solar disk. Instead we only do the averaging
over the central portion of the disk, within $r/r_\odot <0.1$. Inside
this disk-center region there is no significant difference between the
vertical and the line-of-sight directions, so no geometrical
projection is needed to convert the line-of-sight component to a
vertical component. With our previous notations, $B_{\rm app}$ is the
apparent, noise-affected average unsigned vertical flux density within
$r/r_\odot <0.1$, while $B_{\rm ave}$ is its noise-free
counterpart. Like for MDI, Eqs.~(\ref{eq:noisemod}), (\ref{eq:alpha}), and
(\ref{eq:canc}), with cancellation exponent $\kappa=0.13$,  are applicable also
for HMI, since these equations are not resolution dependent
over the spatial scales that we are considering. 

\begin{figure}
\resizebox{\hsize}{!}{\includegraphics{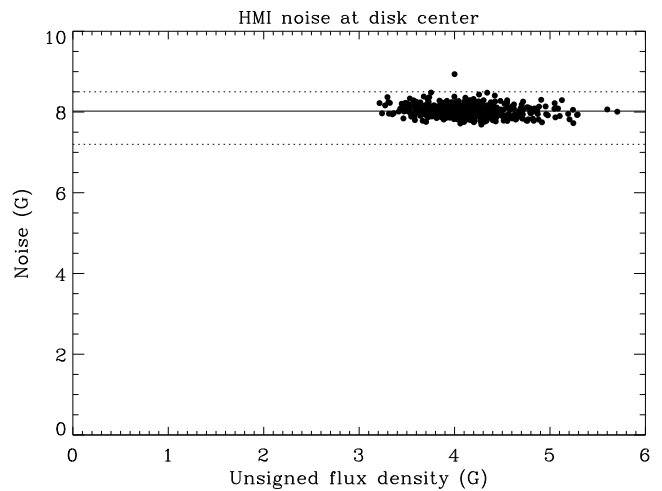}}
\caption{Standard deviation $\sigma$ of the Gaussian noise
  distribution (vertical axis) vs. noise-free average unsigned
  vertical flux density, determined from least-squares fits to the
  disk-center regions of a set of HMI magnetograms, based on the same
  fitting procedure as used for Fig.~\ref{fig:fitmdi}. The average
  $\sigma$ (solid horizontal line) is found to be 8.0\,G. For
  comparison we give as the dotted horizontal lines the upper and
  lower limits to the HMI disk center noise $\sigma$, as determined by
  \citet{stenflo-liuetal12}. 
}\label{fig:hminoise}
\end{figure}

The results of the least squares parameter fitting are shown in
Fig.~\ref{fig:hminoise}, where the noise $\sigma$ is plotted
vs. $B_{\rm solar}$, the value of the noise-corrected $B_{\rm ave}$
for the unsmoothed version of the magnetogram. The average value of
$\sigma$, marked by the horizontal solid line, is found to be
8.0\,G. For comparison we draw as the horizontal dotted lines the
upper and lower limits to the HMI noise as derived from the careful
analysis by \citet{stenflo-liuetal12}. The upper limit of 8.5\,G was obtained by
Gaussian fitting to the core region of the observed PDF in quiet
regions, while the lower limit of 7.2\,G represents the ideal, theoretically
expected level based on Monte Carlo simulations of the photon noise in
HMI. The perfect consistency of our results with these two limits
strongly supports the validity of our noise model based on the use of
scaling laws. 

Note that the HMI $\sigma$ of 8.0\,G is small in comparison with the
disk-averaged $\sigma$ of 18.8 and 33.2\,G for the 5-min and 1-min MDI
magnetograms. Therefore the domination of the noise contribution
relative to the solar basal flux contribution will be much less severe
for HMI as compared with the MDI data set.

\subsection{Basal flux density from the noise-corrected HMI time series}\label{hmitseries}

\begin{figure}
\resizebox{\hsize}{!}{\includegraphics{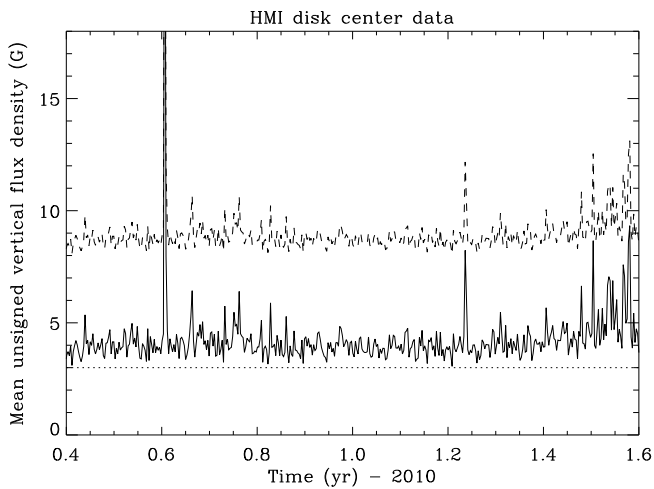}}
\caption{Time series (2010.4 - 2011.6) of the noise-affected average
  unsigned vertical flux density (dashed curve) and its counterpart
  after noise removal (solid curve), representing the disk
  center region ($r/r_\odot <0.1$). The solid curve never dips below
  the 3.0\,G level, which can be interpreted as the basal flux density
  at the HMI resolution of 1\,arcsec. 
}\label{fig:hmiser}
\end{figure}

With the determined value of $\sigma=8.0$\,G we can now, exactly like
we did in Sect.~\ref{sec:rzcorr} for the MDI data, use Eqs.~(\ref{eq:noisemod}) and
  (\ref{eq:alpha}) to convert the noise-affected average unsigned
  vertical flux density $B_{\rm app}$ to its noise-free counterpart
  $B_{\rm solar}$ (which represents $B_{\rm ave}$ for the unsmoothed
  magnetograms). Since we have analysed one HMI magnetogram per day,
  we get a time series, which we have plotted in Fig.~\ref{fig:hmiser}
  for the most quiet period of solar activity, from 2010.4 to
  2011.6. The noise-corrected average flux density $B_{\rm solar}$ is
  plotted as the solid curve, while the time series $B_{\rm app}$
  before noise removal is shown as the dashed curve for comparison. 

While the value of the noise-free  $B_{\rm solar}$ can be seen to vary
significantly from day to day, it never drops below the 3.0\,G level
marked by the dotted line, although the illustrated period represents an
unusually deep and extended quiet phase of the solar activity
cycle. The 3.0\,G level may therefore be interpreted as the basal
unsigned vertical flux density of the Sun, which is always present, 
regardless of the level of solar activity. 

It is difficult to give a good estimate for the error bar in the
  3.0\,G HMI basal flux density level, since the error is most likely
  dominated by systematic effects (like the degree of validity of our
  noise model) rather than by formal fitting errors. In the case of
  the MDI analysis, the difference in the $b_0$ values of 2.7\,G for
  the 5-min integration magnetograms and 1.0\,G for the 1-min
  integration magnetograms gives an indication of the degree of
  significance of the determination, although we give more weight to
  the 5-min magnetograms because they are less noisy to start
  with. Similarly, we give still more weight to the HMI analysis, due
  to the higher quality and much lower noise of the HMI data set, and
  since we only use disk center data with no reference to
  sunspots. 

Still one may wonder if the HMI 3.0\,G level could be much
different, for instance close to zero, due to errors in our
approach. This question will be clarified in the next section by
comparing the appearance of a magnetogram that represents the basal
flux level with one 
that has more than twice as much average flux density.  If the basal
flux density were much less than 3\,G, then the ``basal magnetogram''
would look much more empty, with much less visibility of the magnetic
network, and the flux density histograms for the
two magnetograms would differ much more than they do. From such
considerations we can exclude a basal flux density level below about
2\,G. 

The basal flux level is dependent on spatial resolution (but not
  on integration time, since the temporal evolution of the magnetograms
within the integration times that we are dealing with is
insignificant). The given values refer to
the particular spatial resolution used (1\,arcsec in the case of HMI)
and scales with the resolution according to the law given by
Eq.~(\ref{eq:canc}) with cancellation exponent $\kappa=0.13$. The HMI
3.0\,G level therefore translates to $3.0\times
4^{-0.13}\,=2.5$\,G for the 4\,arcsec MDI resolution. This
scaling is in nearly perfect agreement with the value of 2.7\,G
that we determined from the analysis of the 5-min MDI magnetograms
through the correlation between the unsigned vertical flux density and
the sunspot number over the solar cycle. In the next section we will discuss the
physical meaning of this result.

\section{Basal flux and the local dynamo}\label{sec:locdyn}

\begin{figure}
\resizebox{\hsize}{!}{\includegraphics{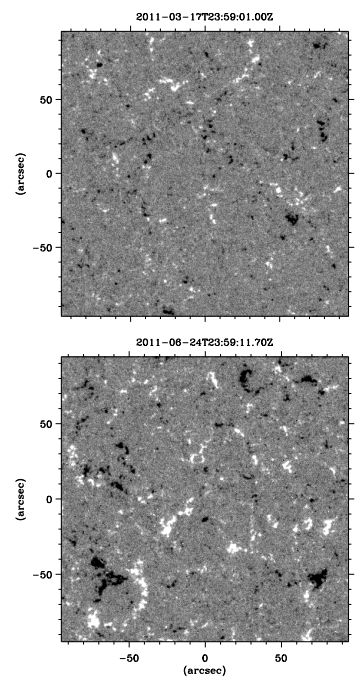}}
\caption{Comparison of the disk-center portions ($\vert x
  \vert,\,\vert y \vert<0.1r_\odot$) of two HMI magnetograms, recorded
  at times 2011.2086 (upper panel) and 2011.4796 (lower panel), when
  the average of the unsigned vertical flux density was 3.06 and
  6.65\,G, respectively. The upper panel thus represents the case when
  the basal flux density level is approximately reached. The grey-scale cuts are at
  $\pm 50$\,G in both panels.  
}\label{fig:hmi2d}
\end{figure}

 While the average unsigned flux density is a convenient parameter to characterize the
statistical properties of the magnetic pattern in terms of a single 
number, we need to examine the full pattern to understand what this
number really means. For this purpose we
illustrate in Fig.~\ref{fig:hmi2d} the disk-center regions of two HMI
magnetograms, both representing the quiet Sun, but with different
values of $B_{\rm solar}$, the noise-corrected average unsigned
vertical flux density. For the upper panel, $B_{\rm solar}=3.06$\,G,
nearly identical with our estimate of the basal flux density level,
while for the lower panel it is 6.65\,G, significantly larger although
still quite small. The plotted regions extend in $x$ (E-W direction)
and $y$ (S-N direction) over $\pm 0.1r_\odot$ with respect to disk
center. The grey scale cuts in both plots are $-50$\,G (dark) and
$+50$\,G (white). The upper and lower magnetograms are from times
2011.2086 and 2011.4796, respectively, and can be identified in the
time series of Fig.~\ref{fig:hmiser} as a local dip and a local peak
at these times. 

While the two magnetograms are qualitatively similar, in both cases
characterized by a magnetic network on the scale of the
supergranulation, they are strikingly different quantitatively,
as expected from the difference of more than a factor of two of their
$B_{\rm solar}$ values. Note that before noise correction the apparent
average unsigned flux densities were 8.1 and 10.8\,G and therefore did not
differ that much from each other (since these values are dominated by the noise
contributions). 

\subsection{Intermittent basal magnetic flux}\label{sec:intermitt}
\begin{figure}
\resizebox{\hsize}{!}{\includegraphics{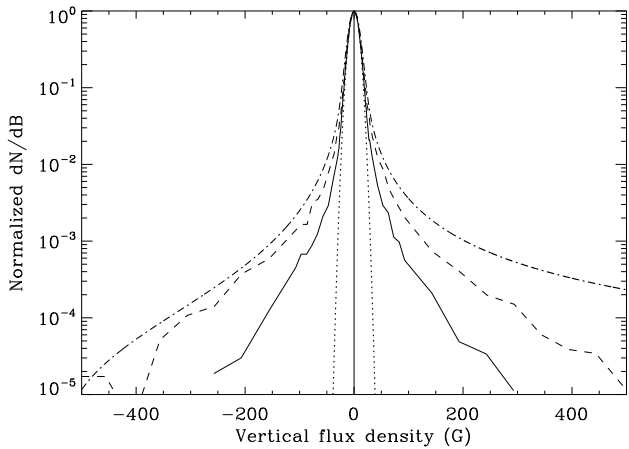}}
\caption{Amplitude-normalized PDFs for the two magnetograms in Fig.~\ref{fig:hmi2d} (solid
  line for the upper magnetogram, dashed line for the lower) and for
  the Gaussian noise distribution with 
  $\sigma=8.0$\,G (dotted line). For comparison we have also plotted as the
  dash-dotted line the analytical representation of the PDF from the
  Hinode quiet-sun analysis of \citet{stenflo-s10aa}, after having
  convolved it with the $\sigma=8.0$\,G Gaussian to make it comparable to the
  HMI PDFs. 
}\label{fig:pdf}
\end{figure}

To make the comparison between the two magnetograms in
Fig.~\ref{fig:hmi2d} more explicit and 
quantitative, we plot in Fig.~\ref{fig:pdf} the respective histograms
of the vertical flux density, as the solid line for the upper magnetogram,
as the dashed line for the lower one. These histograms (which we
  refer to as empirical  probability
density functions PDF, without implying any
assumption concerning the underlying physical process), contain the noise
contribution, since the noise-removal procedure that we applied 
to the average unsigned flux densities can only be used for this
statistical ensemble average. For comparison we have therefore in
Fig.~\ref{fig:pdf} plotted as the dotted curve the Gaussian noise
distribution with the determined standard deviation $\sigma$ of
8.0\,G. It agrees almost perfectly 
in the PDF core region with the two solar PDFs, which only differ from
it by their extended damping wings. 

The effect of noise on the observed PDF is a Gaussian convolution of the intrinsic,
solar PDF. In principle one could therefore retrieve the noise-free solar PDF
through deconvolution. The circumstance that the Gaussian noise
profile almost perfectly matches the core regions of the solar PDFs
means, however, that the core width of the solar PDF is insignificant in
comparison with the noise broadening. 

Through forward modeling with Hinode quiet-sun data from February 27,
2007, it was found that an analytical 
representation in terms of a stretched exponential for the core 
region, supplemented by extended, quadratically declining wings, gave
an excellent representation of 
the observed PDF when convolved with the Gaussian noise distribution
\citep{stenflo-s10aa}. With this analytical
PDF one finds that the average unsigned flux density is almost
exclusively determined by the PDF damping wings, although the PDF core
region represents the vast majority of the pixels 
(the dominating filling factor). In fact, if one 
changes the strength of the wings by decreasing the damping parameter
in the analytical representation, the average unsigned flux density
may go down to values even far below 1\,G, but the exact value of the lower limit
depends on the details of the not so well known intrinsic shape of the
narrow core region. 

It is thus the extended PDF wings that are the source of the determined
(noise-free) values of the average unsigned flux density, including
the value of the basal flux density. The solid-line PDF in Fig.~\ref{fig:pdf} is
representative of the magnetic pattern at the basal flux density
level. The difference in wing amplitude between the dashed and solid
PDFs is fully consistent with the conclusion that the wing region is the source of the
difference between their intrinsic $B_{\rm solar}$ values, 6.65 and
3.06\,G. In contrast, the apparent $B_{\rm app}$ values of 10.8 and 8.2\,G are
dominated by the contribution from the noise-dominated Gaussian PDF
core. 

For comparison with our HMI-based PDFs, we plot
our analytical representation of the mentioned Hinode PDF, but only after it
has been convolved with the Gaussian $\sigma=8.0$\,G PDF. This convolution is
done exclusively for the purpose of allowing us to quantitatively compare the
relative magnitudes of the different PDF wings on a common scale. This
comparison reveals that the quiet region picked by Hinode contained significantly
more magnetic flux than our two HMI regions. The different PDFs in
Fig.~\ref{fig:pdf} illustrate that there is a
broad range of quiet-region flux levels, and that this variable
quiet-sun flux can have nothing to do with a time invariant local dynamo. The
qualitatively similar PDF shapes indicate that all of them are
produced by the same type of physical processes, including the solid-line PDF
that can be taken as representative of the basal flux level. 

The extended PDF wings are a signature of {\it intermittency}, 
  since the combination of high 
  flux density with low occurrence probability in the PDF 
  wings implies significant spatial separation between such relatively
rare, strong-field elements. The
circumstance that the dominant contribution to the unsigned average flux
density always comes from the wings rather than from the PDF core implies that the
magnetic flux is highly intermittent, including the basal flux. The
intermittency expresses itself in Fig.~\ref{fig:hmi2d} in the form of discrete flux
fragments that form a network-like pattern. 

It has been argued that while the PDF wings that represent the
intermittent network-type flux may be asymmetric between
the positive and negative polarities, the PDF core (defined to refer
to the internetwork) is always symmetric, and that
a symmetric, invariant PDF core is a signature of a local dynamo
\citep{stenflo-lites11ssd}.  However, we have demonstrated here that
when the noise contribution is removed from the PDF, the magnitude of
the unsigned average flux density is almost entirely (beyond the 1\,G
level) determined by the intermittent PDF wings. 

The apparent existence of a time invariant basal flux level may seem
to point towards an origin in terms of a local dynamo. The flux
generated by the global dynamo however spreads to
form a background field pattern across the whole surface of the Sun,
and this background does not suddenly vanish when the sunspot number
goes to zero. The decay and removal of flux from the solar photosphere
has a certain time scale, which becomes longer when the flux level
decreases, because flux removal has to do with encounters between
flux fragments of opposite polarities. Due to the extreme flux
intermittency, the polarity encounters become increasingly rare as the
flux density level goes down. Therefore the background flux does not
have enough time to be fully removed before the next cycle injects new flux
into the background. 

This implies that the global dynamo will always leave a
residual, non-zero flux density that will be indistinguishable
from the basal flux density that we have determined. Therefore we can
only say that our basal flux density represents an {\it upper
  limit} to the small-scale flux density that may hypothetically be
generated by a local dynamo.

\subsection{Hidden basal-type flux from the Hanle effect}\label{sec:hidden}
It has long been known from applications of the Hanle effect that the
solar photosphere is seething with a ubiquitous magnetic field that is
tangled on scales much smaller than the telescope resolution and
therefore does not show up in magnetograms that are based on the
Zeeman effect \citep{stenflo-s82,stenflo-s87}. The ``Second Solar
Spectrum'', the linearly polarized spectrum that is exclusively
produced by coherent scattering processes and which is the playground
for the Hanle effect became accessible to systematic exploration in
1994 \citep{stenflo-sk96,stenflo-sk97} through the introduction of the
ZIMPOL technology for high-precision imaging polarimetry
\citep{stenflo-povel95,stenflo-povel01,stenflo-gandetal04}. From the
series of observing campaigns with ZIMPOL since then we have
been able to notice how the general appearance of the Second Solar
Spectrum has changed with the solar cycle, from an ``emission-like''
spectrum (in Stokes $Q/I$) full of intrinsically polarizing lines
during solar minimum, to a mixed emission- and absorption-like
spectrum during the maximum phase \citep[the period extensively documented in the
Atlas of the Second Solar Spectrum,][]{stenflo-gandorf00}. While the
polarization of the molecular lines seemed to be nearly invariant with
the cycle, the atomic lines exhibited large variations
\citep{stenflo-s1spw3}. Another intriguing finding was that the
turbulent field strength derived from the Hanle effect in molecular
lines was much smaller than the value obtained from the Sr\,{\sc i}
4607\,\AA\ line \citep{stenflo-trujetal04}. 

In 2007 a synoptic program was started with ZIMPOL at IRSOL (Istituto Ricerche
Solari Locarno) to explore the solar-cycle variations of the
small-scale, tangled magnetic field with the {\it differential} Hanle
effect \citep{stenflo-setal98} in molecular C$_2$ lines at
5141\,\AA. Assuming a single-valued field
strength and an isotropic angular distribution, an average field of $7.4\pm
0.8$\,G was found for the period 2007-2009, with no evidence for a
significant temporal evolution 
\citep{stenflo-kleint11}. This field therefore appears 
to have the characteristics of a basal flux at the scale where the 
tangled field resides.  

The value of 7.4\,G seems to nicely relate to the 3.0\,G basal level
determined for the HMI spatial scale. Applying the $d^{-\kappa}$ scaling law
with $\kappa=0.13$, we find that 3.0\,G becomes 7.4\,G if we go
down to the 0.7\,km scale. This lies in the middle of the
scale range for the hidden flux 
that was recently estimated \citep{stenflo-s12aa1} based on an
analysis of the magnetic energy spectrum of the quiet Sun over scales 
spanning 7 orders of magnitude. It was argued that
the most logical location for the tangled hidden flux is just below
the range of 10-100\,km, where most of the kG type flux tubes
reside. The tangled fields may exist all the way down to the magnetic
diffusion limit around 10-100\,m, but since their strength is expected
to weaken as we go down in scale, their main contribution should come from
the upper end of the range below a few km. 

In contrast to this modest value for the basal hidden flux 
derived from the differential Hanle effect in the optically thin molecular
C$_2$ lines, the Hanle effect in atomic lines reveals a hidden field that
is an order of magnitude stronger. The most
elaborate modeling so far \citep{stenflo-trujetal04} of the Hanle
depolarization observed in the  Sr\,{\sc i} 4607\,\AA\ line has led to a
field strength of 60\,G for the hidden field (assuming a single-valued
PDF with an isotropic angular
distribution). \citet{stenflo-trujetal04} explain this difference in 
terms of a spatial correlation of the hidden flux with the solar granulation. 
Their modeling shows that the C$_2$
lines are formed exclusively in the interior of the granulation cells, while
the Sr line has contributions from both the intergranular lanes and
cell interiors. The implication is that the hidden, tangled
field is much stronger in the intergranular lanes than in the cell
centers, which appears plausible, since the collapsed kG type flux
also strongly correlates with the intergranular lanes
\citep{stenflo-s11aa}.  The flux of the small-scale tangled field
may therefore be supplied by the decaying flux tubes \citep{stenflo-s12aa1}. 

This interpretation has got support from \citet{stenflo-snik10}, who
could record scattering polarization in molecular CN lines with
sufficient spatial resolution to allow a weak statistical distinction 
between the intergranular lanes and the cell interiors. More Hanle
depolarization was found in the lanes, implying stronger fields
there. \citet{stenflo-shapiro11} did detailed Hanle-effect radiative
transfer modeling of the ultraviolet CN lines and found them to give
field strengths of similar large magnitudes as those found with the
Sr\,{\sc i} 4607\,\AA\ line. The circumstance that the molecular CN lines (and the Sr line)
are optically thick, in contrast to the optically thin C$_2$ lines,
appears to be related to the difference in their behavior. To settle
this issue we need high-resolution mapping of the
Hanle depolarization effect in the Sr line, to determine if the
field-strength contrast between the intergranular lanes and cell interiors
really is as huge as the interpretation models indicate. 

Unfortunately it is difficult to
find suitable atomic line pairs for applications of the differential Hanle effect, in
contrast to many good choices of molecular line pairs. Interpretation of
the Hanle effect in atomic lines therefore depends more on absolute
than differential polarimetry 
as well as on the details of the radiative-transfer modeling,
including the 3-D geometry of the atmosphere, the effects of
the atomic collisions, and the small-scale thermodynamic structure. We need a
differential approach to reduce this model dependence. 

A controversial issue is also whether the strong (of order 60\,G)
hidden field revealed by the atomic lines varies with the solar
cycle. The conspicuous change in the overall appearance of the atomic lines
in the Second Solar Spectrum with the solar cycle is evidence for
substantial cyclic variations of the hidden flux
\citep{stenflo-s1spw3}. In contrast, 
\citet{stenflo-trujetal04} have assembled some evidence from various
observations of the Sr\,{\sc i} 4607\,\AA\ line that the large field
strength that they deduce from this line does not vary with the cycle. We need
a synoptic program for the atomic lines to quantitatively settle this 
issue and determine how high the basal, time-invariant component of
the hidden flux density really is.

\section{Concluding remarks}\label{sec:conc}
The magnetic energy spectrum of the Sun extends over more than 7
orders of magnitude, from the global scales down to the magnetic
diffusion scale around 10-100\,m, where the magnetic field lines
cease to be frozen-in and decouple from the plasma
\citep{stenflo-s12aa1}. Apart from the scales in the range 10-100\,km,
where most of the intermittent flux tubes formed by the convective collapse
mechanism are expected to reside, there is no particular preferred
length scale in this magneto-convective spectrum, which could form a
basis for deciding what is local and what is global. The spectrum
extends all the way down to the magnetic 
diffusion limit, regardless of whether a local dynamo exists or
not. 

The global dynamo is a
source of small-scale magnetic structuring, but it also needs the small
scales for its operation. Much of the large-scale helicity is produced
by the statistical left-right symmetry breaking in the ensemble of small-scale turbulent
elements.  The question whether a local dynamo is physically relevant for the Sun
depends on whether the rate at which it can create new magnetic flux
from a seed field is competitive with the rate at which magnetic flux
is fed into the small scales by the turbulent shredding of flux that
has been created by global dynamo processes. 

Analysis of the observed magnetic flux in MDI and HMI magnetograms
shows that there is an apparently time invariant average unsigned flux
density of 3.0\,G at the 1\,arcsec spatial scale. This basal level
represents an upper limit to the contribution of a hypothetical local
dynamo. It corresponds to 3.5\,G when scaled to the Hinode 0.3\,arcsec
scale with the cancellation function that we
have empirically found to be valid over this scale range. Since the
average unsigned flux densities found in quiet solar regions with
Hinode are much larger, most of the quiet-sun magnetic
flux that has been recorded with  Hinode has nothing to do with a local dynamo. 
 
The average unsigned flux densities are determined almost exclusively by the
extended damping wings of the flux density PDFs, also in the 3.0\,G
case that represents the basal flux. The PDF wings are signatures
of the intermittent nature of the field. A residual background flux
level on the Sun (with a small
number density of intermittent flux fragments) may
survive over relatively long periods in the absence of solar activity
and fresh input from the global dynamo. The 3.0\,G level that we
have found therefore does not necessarily imply a significant contribution
from a local dynamo. An interesting question is
what would happen to the basal-type background flux during a Maunder-type
minimum. If there is a local dynamo, it should maintain a certain flux
density level indefinitely, even if the global dynamo were to be turned off. 

Analysis of the differential Hanle effect in optically thin molecular
C$_2$ lines indicates the existence of a time-invariant, basal-type
field of strength 7.4\,G, representing the region in the interior of
granulation cells. Modeling of the Hanle effect observed in atomic lines (like
the Sr\,{\sc i} 4607\,\AA\ line) and the optically thick CN lines in
the UV gives field strengths that are larger by an order of
magnitude. This apparent contradiction may be explained if the hidden
field is much stronger in the intergranular lanes than in the cell
interiors. It remains controversial how
much of the strong hidden flux represents time-invariant, basal flux. To clarify the
nature of these fields we need to apply differential techniques and map the Hanle effect
with a spatial resolution that resolves the solar granulation. 

If much of the strong  hidden flux (with strength of order 60\,G or
more) really represents a basal flux, then this large field strength cannot be 
directly connected via a fractal scaling law to the 3.0\,G basal 
level that we have determined for the 1\,arcsec spatial scale. While
the 3\,G value places a very tight upper limit to the possible
contribution of a local dynamo to the spatially resolved fields, the
60\,G value, if basal,  offers plenty of room for a 
local dynamo. This leads us to a scenario, in which the relative
contribution of the local dynamo is insignificant at all the scales
that can be resolved by current instruments, but in which the local dynamo
becomes the dominant agent at the smallest scales in the unresolved
domain, probably around and below the 1-10\,km scale \citep[where the hidden flux is
estimated to reside according to][]{stenflo-s12aa1}. The generation of
magnetic energy by the local dynamo at these small scales would then
raise the energy spectrum and flatten it in the domain that leads down
to the magnetic diffusion limit. A synoptic differential Hanle observing program
for photospheric atomic lines is needed to determine whether this
scenario applies to the Sun.

\begin{acknowledgements}
I am grateful for the hospitality of Stanford University during a visit
in March 2012 for work on the SDO/HMI and SOHO/MDI data sets, and for the
help provided by Sasha Kosovichev to develop convenient IDL program
access to the huge HMI and MDI data archives at Stanford. I also want
to thank Yang Liu for helpful discussions about the noise
characteristics of the MDI and HMI magnetograms, and G\"oran
Scharmer for bringing my attention to the effect of noise on 
determinations of the basal flux density. 
\end{acknowledgements}

\end{document}